# The fundamental factor of optical interference

## JiWu Chen[1,*]

[1]*College of Metrology & Measurement Engineering, China Jiliang University, Hangzhou, 310018, China*
*Corresponding author: chen_jiwu@aliyun.com

**It has been widely accepted that electric field alone is the fundamental factor for optical interference, since Wiener's experiments in 1890 proved that the electric field plays such a dominant role. A group of experiments were demonstrated against Wiener's experiments under the condition that the interference fringes made by optical standing waves could have been distinguished from the fringes of equal thickness between the inner surface of emulsion and the plane mirror used to build the optical standing waves. It was found that the Bragg diffraction from the interference fringes formed by the standing waves did not exist. This means optical standing waves did not blacken the photographic emulsion, or the electric field did not play such a dominant role. Therefore, instead of the electric-field energy density solely in proportion to the electric-field square $|E|^2$, "Energy Flux in Interference" was proposed to represent the intensity of optical interference-field and approved in the derivation of equations for the interference. The derived equations indicate that both the electric-field vector and the magnetic-field vector are in phase and have equal amount of energy densities at the interference maxima of two light beams. Thus, the magnetic-field vector acts the same role as the electric-field vector on light interacting with substance. The fundamental factor of optical interference is electromagnetic energy flux densities rather than electric-field alone, or the intensity of optical interference fringes should be the energy flux density of $|\vec{E} \times \vec{H}|$, not electric-field energy density of $|E|^2$.**

## 1 INTRODUCTION

### 1.1 Wiener's Experiment

Up to now, it is widely accepted that electric field alone is the fundamental factor for optical interference. It is because the conclusion was reached that the photochemical action was directly related to the electric vector of light and not to the magnetic after Wiener first experimentally demonstrated the existence of optical standing waves in 1890 [1]. Wiener's experiment setup was: a silvered-plane mirror was illuminated normally by a parallel beam of quasi-monochromatic light to form optical standing waves; and a glass plate coated on one of its plane surfaces with a thin film of transparent photographic emulsion in thickness of less than 1/20 wavelength was placed in front of and inclined to the mirror at a small angle to detect the optical standing waves. After development, the emulsion was found to be blackened in equidistant parallel bands. The blackened maxima corresponded to the intersection between the emulsion and the antinodes of either the electric or the magnetic field. Based on the further experiments, in which the emulsion coated on a glass plate was pressed in contact with a slightly convex mirror, with normal incidence of a monochromatic plane light beam, Wiener concluded that the maxima of blackening were directly related to the electric vector, not to the magnetic vector, because the nodes of the magnetic field were just at the place of the antinodes of the electric field according to electromagnetic theory. In 1891, Lippmann first applied it in color-photography bearing his name on the basis of optical standing waves [2].

Subsequently, similar experiments were performed by Drude and Nernst in 1892 using fluorescent films [3] and by Ives and Fry in 1933 using photo-emissive films [4] respectively as the detectors instead of the emulsion that was used by Wiener to sense the light. These experiments and the conclusion had become the acceptable knowledge and were also introduced in a classic optics textbook by M. Born and E. Wolf, who explained: "the electromagnetic force on a charged particle at rest is proportional to the electric vector" [5], namely, the electric field plays such a dominant role.

In recent years, Minoru Sasaki et al. in 1999 [6], H. Stiebig and H-J Büchner et al. in 2003 [7-9] built standing-wave interferometers using photodiodes thinner than optical wavelength as the detectors of standing waves and described the characters of the interferometers according to the conclusion drawn by Wiener.

### 1.2 The Problems in Wiener's Experiment

In those theoretical analyses for any case of the experiments above, the films as the detectors were regarded as being immersed in the optical standing waves. However, taking Wiener's experiment as an example, on the emulsion still appeared the interference fringes of equal thickness made by the two light beams reflected from the plane mirror and the inner surface of the emulsion. And the fringes were just overlapped on the fringes of standing waves in the thin emulsion film. Therefore, the severe problem in Wiener's experiment was that the standing-wave's interference fringes on the emulsion could not be distinguished from ones of equal thickness when the emulsion was as thin as 1/20 wavelength. So the conclusion originated from it might be problematic, or even wrong. The interference fringes of equal thickness were similar to Newton's rings or the fringes on the soap film being illuminated by a parallel light beam. The theoretical proof is followed below.

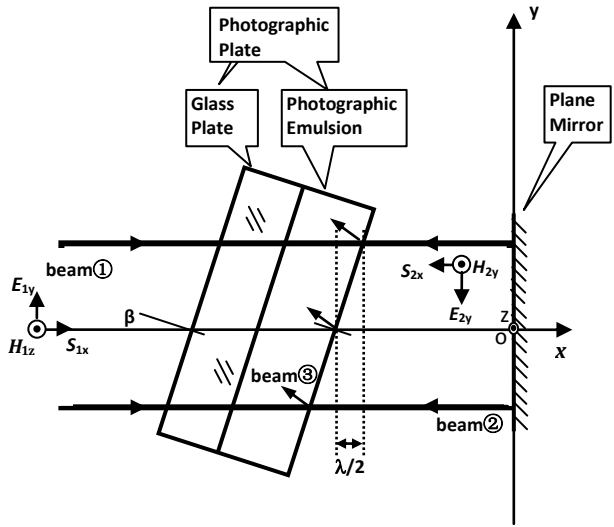

**Fig.1. The interference fringes in the Wiener's experiments (the photographic plate is simplified as an equivalent air-layer and the thickness of the emulsion is greatly exaggerated.)**

As shown in Fig.1, a parallel light beam Beam ① with electric-field $E_{1y}$ and magnetic-field $H_{1z}$ normally illuminates a plane mirror. The reflected beam Beam ② was with $E_{2y}$ in direction opposite to $E_{1y}$ for its having π out-of-phase on the mirror, and $H_{2z}$. If there is no any loss, the electric-field $E_{1y}$ and $E_{2y}$ could be expressed by

$$E_{1y} = a \cdot \exp(ikx) \quad (1)$$

$$E_{2y} = -a \cdot \exp(-ikx) \quad (2)$$

The superposition of these two electric-fields is

$$E_y = E_{1y} + E_{2y} = 2ai \cdot \sin(kx) \quad (3)$$

The time-averaged electric-field energy density of the superposition is

$$\left\langle D_{E_y} \right\rangle_T = \frac{\varepsilon |E_y|^2}{4} = \frac{\varepsilon \cdot a^2}{2}[1 - \cos(2kx)] \quad (4)$$

where, the energy density varies along x-axis, with antinode's spacing of $\lambda/2$. The energy density equals zero on the mirror.

As the photographic plate was inclined at the angle $\beta$ from the mirror, the traces with the spacing of $\lambda/2\sin\beta$ would be left on the surface of the emulsion by the standing waves. It was just predicted by Wiener as a result from Wiener's experiment.

However, there were other interference fringes in it. Beam ① could be reflected by the inner surface of the emulsion to form its reflection Beam ③. Beam ③ could interfere with Beam ② from the mirror. Their interference fringes lay along the bisector between the two beams with the same fringe spacing of $\lambda/2\sin\beta$ as the standing waves did. If one point at the edge of the emulsion touched the mirror, the interference would be a minimum at the touched point. It was because that Beam ③ reflected by the emulsion had no π out-of-phase even if the thin thickness of the emulsion was less than 30nm while Beam ② had π out-of-phase. So the interference fringes formed by Beam ③ and Beam ② left the traces with the same spacing and the same beginning point as the standing waves did. That is, the interference fringes would have the same traces as by the standing waves in the thin emulsion with the thickness of less than 30nm.

On the other hand, we should not ignore the interference fringes formed by the Beam ③ and Beam ② though the intensity I (Beam ③) was very small. In fact, the element of the interference with Beam ② was its electric field E (Beam ③), rather than the intensity of Beam ③. Let us estimate the visibility of the fringes made by the interference between Beam ③ and Beam ②. Suppose that the reflectance of the mirror was 1, and the reflectance of the Photographic Emulsion, just similar to a glass surface, was 0.04. Then I (Beam ②) = 25*I (Beam ③). But E (Beam ③) = 0.2 (= $\sqrt{0.04}$ ) and E (Beam ②) = 1. Therefore, the fringe's visibility ≅ 0.4. It was not very small. A simple experiment was demonstrated in UNNECESSARY APPENDIX a.

In other words, the traces by the standing waves could not be distinguished from the interference fringes formed by Beam ③ and Beam ② in Wiener's experiment. In fact, it was Beam ③ & ②' fringes that appeared in Wiener's experiment, because no traces by the standing waves existed as explained below and shown up in my experiments.

## 2 THE IDEAL EXPERIMENTS AND THE DESIGN OUTLINE OF MY EXPERIMENTS

### 2.1 The Ideal Experiments Indicating No Standing-wave's Interference Fringes

One might be convinced that there are no interference fringes in the emulsion formed by standing waves by going through the following two ideal experiments.

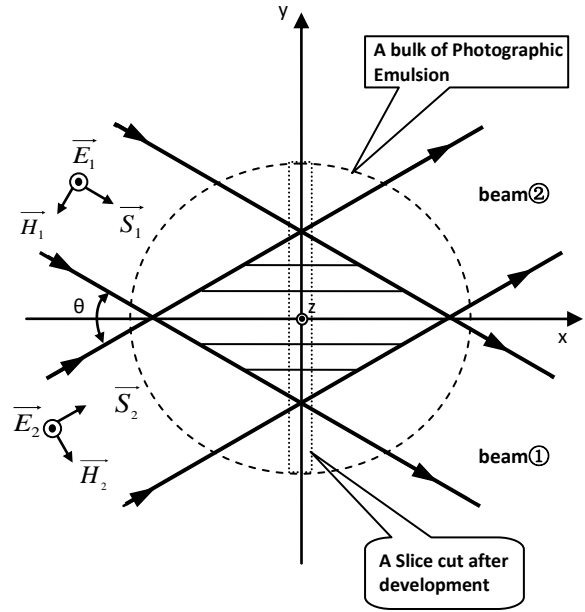

**Fig.2. The process about two-beams' interference fringes becoming to the fringes made by a standing wave**

**The first ideal experiment**

Suppose there are two light beams, with the same magnitude of electric-field polarized linearly perpendicularly to the paper plane, intersecting each other with angle $\theta$ in the intersecting area, and a bulk of photographic emulsion placed in the interference zone, as shown in Fig.2. After exposure and development, cut off a thin slice from the emulsion bulk, whose surface is perpendicular to the bisector of the beams, named by "Slice $\theta$". Then, angle $\theta$ is increased by $\Delta\theta$ to ($\theta+\Delta\theta$), and another bulk of photographic emulsion is placed in the interference zone. Similarly, after exposure and development, cut off another thin slice, named by "Slice ($\theta+\Delta\theta$)". Comparing the two slices, we can find: (1) the interference fringes are denser on "Slice ($\theta+\Delta\theta$)" than those on "Slice $\theta$" as the fringe spacing is equal to $0.5\lambda/\sin(\theta/2)$; (2) the magnitude of blackening is weaker in "Slice ($\theta+\Delta\theta$)" than "Slice $\theta$" as the lux on the Slice is proportional to $\cos(\theta/2)$. By noticing that the interference fringes in the emulsion bulks were parallel to the bisector of the beams, it should be concluded that every interference fringe in "Bulk ($\theta+\Delta\theta$)" is weaker than in "Bulk $\theta$". While increasing angle $\theta$, the interference fringes in the emulsion bulk become weaker by the ratio of $\cos(\theta/2)$. Once $\theta$ turns to 180°, the interference fringes will become null. That means there are not any interference fringes in standing waves, or the electric field did not play such a dominant role.

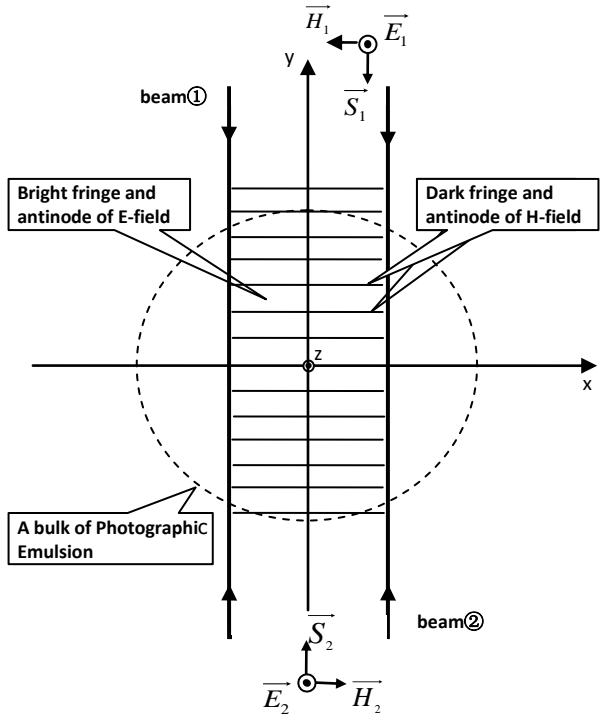

**Fig.3. The counterevidence to show that standing waves could not make any fringes in the emulsion bulk**

**The second ideal experiment**

Let me give another explanation to the same conclusion by counterevidence, as shown in Fig.3. Assuming that standing waves could make interference fringes in the emulsion bulk, the direction of the bright and dark fringes were perpendicular to the optical axis along which the two light beams would propagate oppositely. But the fact was that no optical energy could pass the dark fringes, otherwise the dark would become bright. A bulk of photographic emulsion was placed in the standing waves zone. The exposure needed time to complete. Once a molecule of emulsion was exposed in the electric-field maximum of the standing waves, it must absorb some electric energy, then, the standing waves needed more energy to hold on its former state. Thus, it was the matter: where would the optical energy come from? It would have to supply more energy to the standing waves passing through the dark fringes, and made them "bright". It also means that there are never any interference fringes in standing waves.

**2.2 The Experiments Distinguished Between the Interference Fringes of Standing Waves and Equal Thickness**

In this paper, a set of experiments on the basis of exact optical standing waves are demonstrated. A completely different conclusion has come out against Wiener's.

The design of the experiments was based on the different directions between the interference fringes of standing waves and equal thickness. A general kind of holographic recording medium with the emulsion's thickness of much thicker than 30nm was chose as the Photographic Plates. When a thin laser beam in the original incidence path illuminated on a developed photograph placed in the same optical path as the photograph was exposed, the Bragg diffraction from Bragg grating, formed in the emulsion by the interference fringes of standing waves, should be in the anti-incidence direction while the diffraction from the interference fringes of equal thickness should transmit through the photograph and form in the two directions at the doubled incidence angle of the photograph.

Only when the two-beams intersecting angle $\theta$ being exactly 180° with the precision of Arc Seconds, the phenomenon without standing-waves' interference fringes could happen as mentioned above. The experiments were not easy to do because the efficiency of the diffraction from Bragg grating was very high. Even a very weak Bragg grating, formed in the emulsion by the interference fringes with angle $\theta$ of very close to 180°, could make Bragg diffraction be found.

If the two flat surfaces of the photographic plate were not exactly parallel to each other, once a photographic plate was placed in the optical path for exposure, the surfaces behaving like a glass wedge would deviate the light beam which passed through the photographic plate. So for exact optical standing waves, the problems required to solve include: (1) to avoid the photographic plates, usually like glass wedges, disturbing the standing waves; (2) to minimize the aberration of the parallel light beams.

If the incidence angle was set to the Brewster's angle of the emulsion to achieve no reflection from the emulsion, there must be only the interference fringes of standing waves in the developed photograph. In this case, when the developed photograph was tested with a thin laser beam, the existence of interference fringes of standing waves was proved if Bragg diffraction was found. Otherwise, the interference fringes of standing waves did not exist.

In this paper, the formulations for describing optical interference are also derived by means of electromagnetic energy flux densities.

## 3 EXPERIMENTS ON THE BASIS OF EXACT OPTICAL STANDING WAVES

### 3.1 Setup

The arrangement for the experiments is illustrated in Fig.4. The Laser Beam from Newport Model: R-30991 (HeNe Laser, 633 nm, 5.0 mW, 500:1 Polarization, Longitudinal Mode: 441 MHz) passing through the Polarizer with polarization direction in the plane of incidence or the paper of Fig.4, was focused by the Microscope Objective Lens at the Pinhole with an aperture of 5μm, and then was expanded by Telescope Objective Lens1 with a focal length of 365mm and an aperture of 65mm to form a parallel beam with a diameter of 65mm. Only a part of the parallel beam was used by an aperture of 25mm to form optical standing waves and was recorded on the photographic emulsion. The parallel beam with 25mm diameter through the Glass Plate coated with photographic emulsion passed through Telescope Objective Lens2 identical to Telescope Objective Lens1 and focused at the Plane Mirror surface normal to the optical axis. The reflected light from the Plane Mirror traveled back in its coming way and superimposed onto the incident light beam to form the optical standing waves in the emulsion which could show the profile of the standing waves after development.

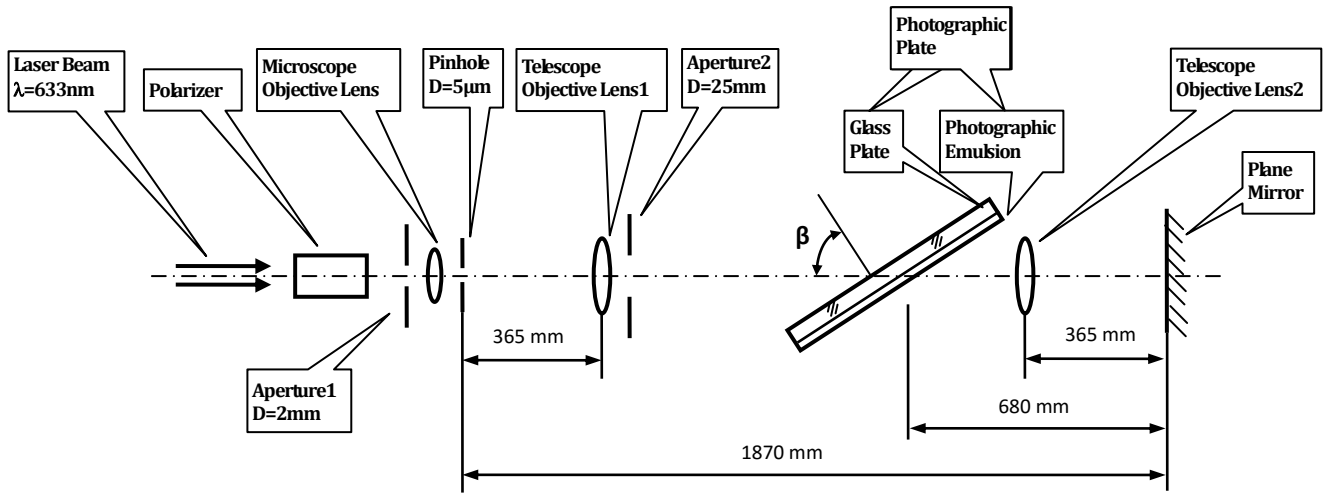

**Fig.4. The setup for experiments on exact optical standing waves (plane mirrors only for beam turning were not illustrated)**

Due to the existence of multiple longitudinal modes in the HeNe Laser, two laser beams can interfere well with each other when their optical path difference equals to even multiple of the laser cavity length, which shows identical phenomenon of optical interference as appeared when the path difference is zero; and the two beams cannot interfere with each other when their optical path difference equals to odd multiple of the laser cavity length. According to this principle, the optical path of 2×1870mm between the Pinhole and Plane Mirror was set to equal to 11× the laser cavity length of 340mm, so that the light reflected by the Pinhole on which the Pinhole was imaged again by the light coming back from the Plane Mirror could not interfere with the incidence. The optical path of 2×680mm between the Photographic Plate and Plane Mirror was set equal to 4× the laser cavity length, so that the incidence and the light reflected from the Plane Mirror could interfere well with each other.

Telescope Objective Lens2 was the key element different from ordinary scheme to generate the standing waves. Because the surface of the emulsion was hardly exactly parallel to the glass surface of the Photographic Plate, the whole Photographic Plate was like a glass wedge to deviate the light beams passing through it. When the surfaces of the emulsion and the glass side of the Photographic Plate were exactly planar, Telescope Objective Lens2 could warrant the reflected light beam to be just in the opposite direction of the incidence. The details are shown in UNNECESSARY APPENDIX b.

**3.2 Adjustment**

It needs accurate adjustments to build standing waves. The accurate adjustments can ensure exact parallel light beams without optical aberration exactly in the opposite directions in which the incident and reflected light propagate respectively.

Several adjustment steps were applied to making optical parts be in exact coaxiality, and the Photographic Plates made from optical glass were used [APPENDIX A], for the least optical aberration in the optical standing waves.

**3.3 Results and Analysis**

Light intensity of 0.6~1.0μW/mm2 was detected by a light power detector at the center of the expanded parallel laser beam directly from Telescope Objective Lens1. The air condition was shut off during the experiments to minimize the vibration of the work table. Subsequently, the fluctuation of light intensity caused by slightly unstable ambient temperature in the lab could be compensated by the light exposure that was the integration of light intensity over exposure time. The exposure time was controlled by an electric shutter in order to ensure the amount of exposure within 0.13~0.64 μJoule/mm2 and the optical density of the photographs within 0.2~2.9 after development, in the linearity region of the emulsion. The incidence angle, angle β (referring to Fig.4), between the surface normal of the Photographic Plate and the optical axis, was respectively set to 57.5° and 2.5° for two groups of the experiments. 57.5° was the Brewster's angle of the emulsion, very close to the Brewster's angle of the Glass Plate. Eight photographs were exposed on every Photographic Plate with a rectangular shape of 90mm×240mm at the incidence angle of 57.5°, and sixteen photographs at 2.5°. In the 2.5° incidence angle's experiments, a piece of black paper was placed at a proper position before the Plane Mirror in Fig.4 for eliminating the light beam spot from the Photographic Plate reflecting the light coming from the Plane Mirror. The black paper was not illustrated in Fig.4.

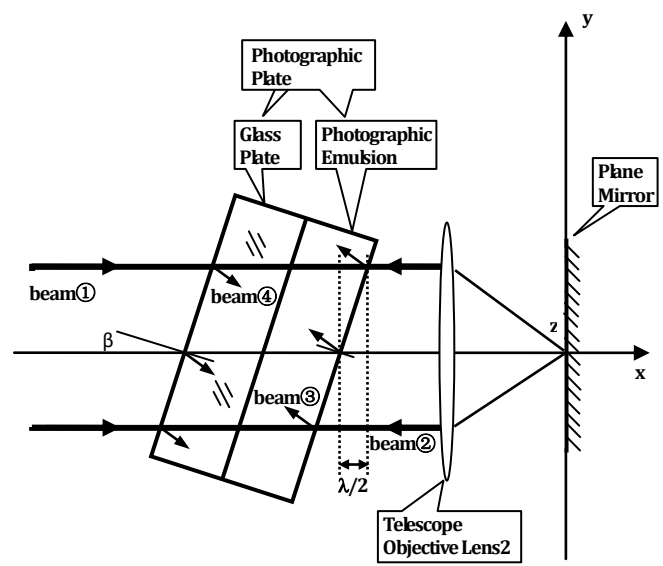

**Fig.5. The partial setup in my experiments (the photographic plate is simplified as an equivalent air-layer and the thickness of the emulsion is greatly exaggerated.)**

In Fig.5, besides Beam ①, Beam ② and Beam ③, there was another reflected beam, Beam ④, which was the reflected beam from the glass inner surface of the Photographic Plate on which Beam ② irradiated. For an easy explanation, hereafter some other descriptions are symbolized. The fringes interfered between Beam ④ and Beam ② was Fringe ②④. In this way, six interference patterns were Fringe ①②, Fringe ①③, Fringe ①④, Fringe ②③, Fringe ②④, and Fringe③④. Fringe ①② was formed by the standing waves if existed. Fringe ①③, Fringe ①④, Fringe ②④, and Fringe③④ did not show up in Wiener's experiment because of their optical path difference beyond the coherence length of Wiener's light source.

The following measurements for a developed Photographic Plate are needed. (i) the diffraction from a set of fringes. If the diffraction was able to be detected, the corresponding fringes would exist. (ii) The flatness of the emulsion surface. A high level flatness could ensure Telescope Objective Lens2 working well.

(i) To detect diffraction from a developed Photographic Plate. I placed the developed Photographic Plate in the same mount as exposing the emulsion, a light Power Detector near the optical axis to get the intensity of the Bragg diffraction and a white paper before the Plane Mirror as a screen in case of reflection from the Plane Mirror. Only a thin laser beam was projected on the center of the developed photographs after removing the Microscope Objective Lens and Pinhole and Telescope Objective Lens2, shown by dotted lines in Fig.6, from the optical path. When placing the light Power Detector in the path of transmitted diffraction at the angle of 2β, as shown in Fig.6, obtained could be the intensity of the ±1st order diffraction from Fringe ①④ and Fringe ②③.

(ii) To measure the flatness of the emulsion surface. By means of illuminating every photograph by an expanded parallel light beam, the flatness and parallelity of the two surfaces of the Photographic Plate could be measured according to the fringes of equal thickness. In general, photograph's parallelity could be figured out by the number of the equal-thickness fringes. The less the number was, the better the parallelity was. Emulsion's flatness could be estimated by the distribution of the equal-thickness fringes. The more uniform the distribution was, the better the flatness was. The number of the equal-thickness fringes and their distribution are also listed in the summarized table.

**(1) For the photographs when exposed at the incidence angle of 57.5°**

There was no reflection from the surfaces of the Photographic Plates because 57.5° was the Brewster's angle of the emulsion and the Glass Plate. The only one interference pattern of Fringe ①② could be found in the detection of a developed Photographic Plate as in Fig.6. The pattern was a group of interference fringes parallel to each other and vertical to the optical axis, called Bragg grating or the standing wave pattern. When a laser beam illuminated the Bragg grating along the optical axis, it could cause Bragg diffraction going back in the same way as the incident light. When we turned the developed Photographic Plate slightly, the Bragg diffraction could be found near the axis, and even detected by the Power Detector as in Fig.6.

The diffraction resulted from the developed photographs at 57.5° under a thin laser beam with a power of about 3mW are summarized in Table 1. The number of the equal-thickness fringes and their distribution are also listed in the table to show the parallelity and flatness of the surfaces at the two sides of Photographic Plate. Most of the photographs were exposed under the condition mentioned above. Among the 72 photographs, 4 photographs were illuminated by an expanded laser beam with a diameter of 55mm and a Pinhole with a clear aperture of 10μm was used while exposing them.

**Table 1. Statistics of the results of photographs exposed at 57.5° and their parallelity & flatness listed as the equal-thickness fringes' number & distribution**

| | Number of photos | Fringes' Number & Distribution |
|---|---|---|
| No Bragg diffraction | 11 | was less than 4, or uniformly |
| Bragg diffraction: 4~20nW | 17 | about 4 and disorderly |
| Bragg diffraction: 20~445nW | 44 | much more than 4, and disorderly |
| Diffraction of the ±1st order | 0 | / |
| Total of the photographs | 72 | |

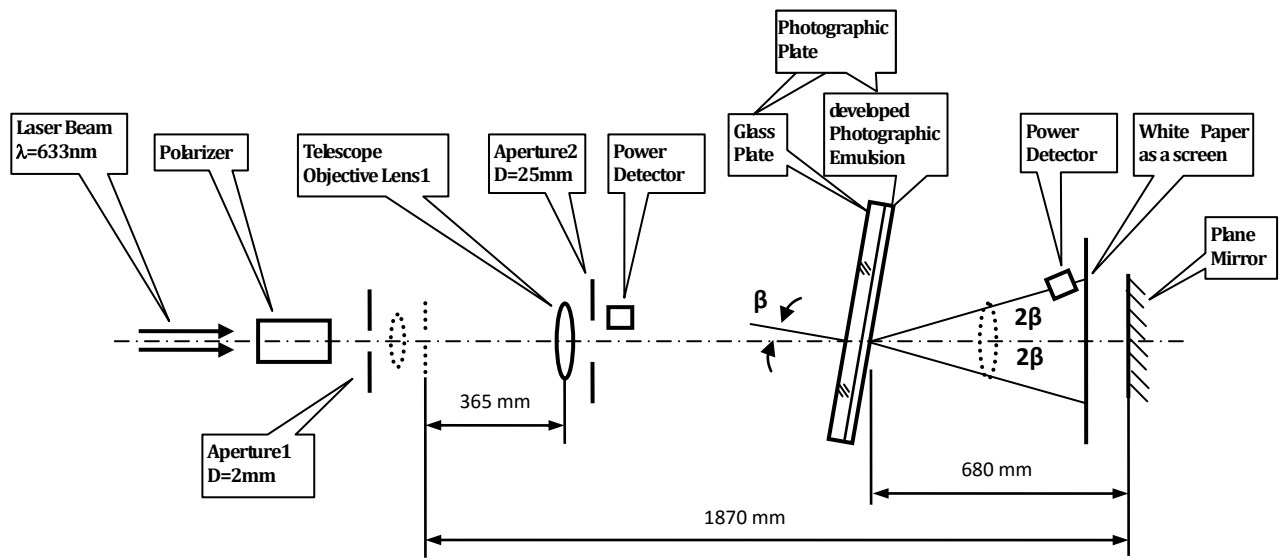

**Fig.6. The detection of a developed photographic plate (plane mirrors only for beam turning were not illustrated and the dotted ones were removed out)**

According to Table 1, the results for the Brewster's angle of 57.5° are analyzed below:

i. The diffraction of the -1st order could not be found on any of the photographs because its corresponding reflection was absent at Brewster's angle 57.5° from the Photographic Plates' surfaces, so that there were no any interference patterns except Fringe ①② in the photographs. The angle of 57.5° was too big to have the +1st order diffraction on the photographs.

ii. Bragg diffraction was not found on 11 photographs, with the phenomenon that the number of the equal-thickness fringes on the photographs was less than 4 or the fringes distributed uniformly. It indicated that every photograph without Bragg diffraction was a perfect glass wedge formed by two flat surfaces, the glass surface and the emulsion surface of the Photographic Plates. It is only in this case that Telescope Objective Lens2 in Fig.4 and Fig.5 could make the incident light beam exactly go back in the coming way to form exact standing waves.

In the earlier standing wave experiments, none of photographs without Bragg diffraction was found if all the optical parts were not precisely adjusted to reach the exact coaxiality. If Telescope Objective Lens2 was not placed as in Fig.4 or the Photographic Plate made of optical glass was not used, the number of the photographs without Bragg diffraction went down sharply. Therefore in order to warrant the success of the experiments, it was required to place Telescope Objective Lens2 on the optical path and to use the Photographic Plate made of optical glass, and to pay special attention to eliminating the optical aberration brought by inexact coaxiality.

As a result of the experiments, it was concluded that there was no Bragg diffraction on the photographs exposed in the exact optical standing waves, namely, there were not any interference fringes in standing waves, or the electric field did not play such a dominant role.

iii. Bragg diffraction was found on most of the photographs, with the phenomenon that the number of the equal-thickness fringes on those photographs was more than 4 and the fringes were distributed disorderly. That meant the reflected light from the Plane Mirror was not an exact plane wave or not traveling back just opposite to the incident light beam. Thus the standing waves were not exactly built in the emulsion. The reflected light interfered with the incident light to form the interference fringes in the direction almost perpendicular to the optical axis. The Bragg diffraction was built by the interference fringes instead of the standing-waves' fringes.

iv. Bragg diffraction with very weak light power of less than 20nW was found on 17 photographs and the equal-thickness' fringes distributed disorderly, though the number of the fringes was about 4. It indicated that Bragg diffraction varied continuously with the angle between Beam ① and Beam ②.

Though Telescope Objective Lens2 was introduced into the setup referring to Fig.4 and Fig.5, to significantly improve to my experiments, the problem of using the lens was that the rays at the up-side of the coming beam would flip to the down-side of the returning beam. Namely, the returning wave transect would turn 180° against the coming wave. Therefore, Telescope Objective Lens2 could only function well when the two surfaces of the Photographic Plate were very flat. It could not function as designed if there were some defects on local areas of the Photographic Plate.

**(2) For the photographs when exposed at the incidence angle of 2.5°**

Since there were six interference patterns on the photographs in this case, the phenomenon of the diffractions from the interference patterns was much more complicated than at 57.5°.

The ±1st order diffraction from Fringe ②③ showed up definitely but was never mentioned in Wiener's experiment. Actually it was Fringe ②③ that appeared in Wiener's experiments.

The conclusion could also be drawn that the optical standing waves were not able to make any interference fringes in the emulsion, according to the results of the developed photographs exposed at 2.5°. The details are presented in APPENDIX B.

## 4 DERIVATION FOR ENERGY FLUX IN INTERFERENCE FIELD

The well known 'Poynting vector' is the source of the definition of light intensity. Up to now, it has not been applied to the analysis of optical interference because it has commonly been accepted that the fringes exist in standing waves. So it might be better to review Poynting vector before the derivation for energy flux in interference field.

The energy flux density or Poynting vector for a parallel light beam is $\vec{S} = \vec{E} \times \vec{H}$. According to the electromagnetic equation, $\sqrt{\varepsilon}|E| = \sqrt{\mu}|H|$, its time-averaged value of energy flux density is

$$\left\langle \left|\vec{S}\right| \right\rangle_T = \frac{1}{2}|E|\cdot|H| = \frac{1}{2}\sqrt{\frac{\varepsilon}{\mu}}|E|^2 = \left\langle w_{em} \right\rangle_T \cdot V$$

where $V = 1/\sqrt{\varepsilon\mu}$ is the light velocity, and $w_{em} = \frac{\varepsilon}{2}|E|^2$ is its energy density of electromagnetic-field.

If the constant coefficients, $\frac{1}{2}\sqrt{\frac{\varepsilon}{\mu}}$, are omitted, the intensity of the beam or energy flux density is determined only by the square of its electric field, $|E|^2$.

According to Wiener's conclusion, $|E|^2$ alone has been widely accepted to represent the intensities of light's beam and interference fringes in optical interference-field.

But $|E|^2$ is electric-field energy density in interference-field, not energy flux density. This becomes obvious in standing waves where the energy flux density is zero.

Energy flux density should be reconverted to represent the intensity of interference fringes in optical interference-field. The following will show that interference-field is actually the superposition of the standing waves and the interference-energy-flux in the area of two-beam's intersection. And only the interference-energy-flux can make the fringes rather than the standing waves.

By resolving the energy flux densities or Poynting vectors $\vec{S_1}$ and $\vec{S_2}$ into the components, $S_{1x}$ and $S_{2x}$ in the bisector direction of the two beams or x-axis, and $S_{1y}$ and $S_{2y}$ in y-axis respectively, as shown in Fig.7, we can get:

$$\vec{S_1} = S_{1x}\vec{x_0} + S_{1y}\vec{y_0} \text{ and } \vec{S_2} = S_{2x}\vec{x_0} + S_{2y}\vec{y_0}$$

As $\vec{E_1} = E_{1z}\vec{z_0}$ and $\vec{E_2} = E_{2z}\vec{z_0}$, the resolution of Poynting vectors corresponds to resolving only the magnetic field vectors into the components:

$$\vec{H_1} = H_{1x}\vec{x_0} + H_{1y}\vec{y_0} \text{ and } \vec{H_2} = H_{2x}\vec{x_0} + H_{2y}\vec{y_0}$$

Then in the intersecting area of the beams, the superposition of the magnetic-fields and electric-field are,

$$H_x = H_{1x} + H_{2x} \text{ and } H_y = H_{1y} + H_{2y}, \text{ and}$$

$$E_z = E_{1z} + E_{2z}.$$

Thus, the instantaneous energy flux density along y-axis is,

$S_y = E_z \cdot H_x$, then if the electric-field's and magnetic-field's amplitudes are *a* and *b*, respectively,

$$S_y = ab\sin\frac{\theta}{2}\sin(2ky\sin\frac{\theta}{2}) \cdot \sin[2(\omega t - kx\cos\frac{\theta}{2})] \quad (5)$$

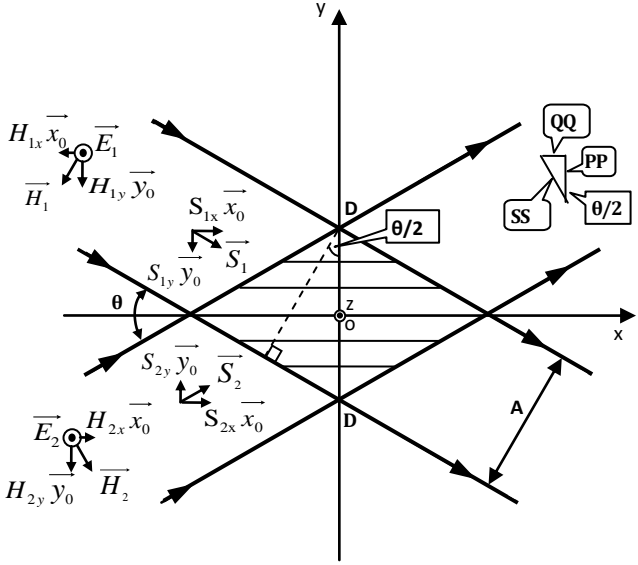

**Fig.7. The flux of energy in interference fields and standing wave fields**

It indicates existence of the standing-wave component in the interference field, as the magnetic-field component $H_x$ is in quadrature with the electric-field vector $E_z$.

Then its time-averaged value of the energy flux density along y-axis is

$$\langle S_y \rangle_T = 0 \quad (6)$$

Though the component along y-axis is $S_y = S_{1y} + S_{2y} = 0$ apparently at x-axis as shown in Fig.7 and the expression's meaning could be understood in the way that the electromagnetic fluxes traveling oppositely along y-axis were equal to each other, Equation (6) actually presented its physical meaning which is just the result of the standing waves of the component. In the other words, no electromagnetic flux travels along y-axis, or passes interference fringes.

The corresponding instantaneous energy flux density through yz-plane along x-axis, called Interference Energy Flux Density, is,

$S_x = -E_z \cdot H_y$, then

$$S_x = 4ab\cos\frac{\theta}{2}\cos^2(ky\sin\frac{\theta}{2}) \cdot \cos^2(\omega t - kx\cos\frac{\theta}{2}) \quad (7)$$

where, the component $H_y$ of the magnetic-field is in phase with the electric-field vector $E_z$. The time-averaged value of the energy flux density along x-axis is,

$$\langle S_x \rangle_T = 2ab\cos\frac{\theta}{2}\cos^2(ky\sin\frac{\theta}{2}) \quad \text{, or}$$

$$\langle S_x \rangle_T = ab\cos\frac{\theta}{2}[1 + \cos(2ky\sin\frac{\theta}{2})] \quad (8)$$

or

$$\langle S_x \rangle_T = \sqrt{\frac{\varepsilon}{\mu}}a^2\cos\frac{\theta}{2}[1 + \cos(2ky\sin\frac{\theta}{2})] \quad (9)$$

$S_x = S_{1x} + S_{2x}$ is the interference energy flux density according to Fig.7. The interference result of the two parallel light beams is shown by Equation (8) or (9). It misses the ratio of cos(θ/2) in the general equation derived only from the electric-fields. Equation (8) or (9) is the distribution of the energy flux density in the interference field including electric and magnetic field rather than electric-field energy density alone.

According to Poynting's theorem, it can be proved that in the interference energy flux density or the standing-waves, both the electric energy density and the magnetic energy density are equal to each other. It consists with the wave principle that electromagnetic waves can only exist as the electric energy and magnetic energy are converting to each other.

The details are presented in APPENDIX C.

### Explanation of the First of Ideal Experiments

Equation (9) is the time-averaged energy flux density for forming interference fringes. If the constant coefficients of $\frac{1}{2}\sqrt{\frac{\varepsilon}{\mu}}$ are omitted, it becomes

$$\langle S_x \rangle_T = 2a^2\cos\frac{\theta}{2}[1 + \cos(2ky\sin\frac{\theta}{2})] \cdot$$

It is identical to the interference pattern only by means of electric-fields except the coefficient of $\cos(\theta/2)$.

The average electromagnetic density in the intersecting area is $2a^2$, the sum of electromagnetic energy densities of the two beams. It consists of $2a^2\sin^2(\theta/2)$ for the standing waves and $2a^2\cos^2(\theta/2)$ for the interference-energy-flux traveling through the yz-plane along x-axis. The electromagnetic energy density $2a^2\cos^2(\theta/2)$ for the interference field corresponds to the energy flux density $2a^2\cos(\theta/2)$ in the direction of x-axis.

### Interference Fringes in Standing Wave

It is the coefficient of $\cos(\theta/2)$ in Equation (8) or (9) that makes things different. If the angle $\theta$ between the two parallel beams increases, the average energy flux density $\langle S_x \rangle_T$ for interference fringes must decrease. Once the angle $\theta$ equals 180° just as in the case of standing waves, no interference fringes will show up because the average energy flux density $\langle S_x \rangle_T$ turns to zero. Therefore, standing waves were not able to make any interference fringes in emulsion.

## 5 CONCLUSION

The experiments have been demonstrated under the condition that the interference fringes made by optical standing waves have been distinguished from the fringes of equal thickness between the inner surface of emulsion and the plane mirror. Because the Bragg diffraction from the Bragg grating formed by the standing waves was not found, the conclusion has been drawn that optical standing waves did not blacken the photographic emulsion. Namely, there were not any interference fringes in standing waves, or the electric field did not play such a dominant role.

In fact, optical standing waves are only a form of energy storages. They behave as electromagnetic energy oscillates in an inductance and a capacitor with the attribute of the electric current and voltage being in quadrature. No energy streams out of any standing waves, because both the electric-field vector and the magnetic-field vector being in quadrature with each other leads the averaged energy flux density or averaged Poynting vector of optical standing waves to be zero. Especially on the maxima or antinodes of the electric-field, the instantaneous energy flux density of standing waves is also zero, as the antinodes of the electric-field are just at the nodes of the magnetic-field in the standing waves. It is obvious that no substance including emulsion can be changed without supplying energy continuously or applying energy flux on it. In the other words, there are never interference fringes of standing waves in emulsion. Once the energy storage was absorbed by something, it must be in the way of electromagnetic energy flux with the attribute of the electric-field and magnetic-field being in phase, just like a lamp being lighted by electric current and voltage being in phase, in which the electric current corresponds to the magnetic-field and the electric voltage corresponds to the electric-field.

Photons instead of waves can interact with substance, which is a conclusion in quantum mechanics. If one asked which character of electromagnetic waves corresponds to photons, the answer should be the energy flux, because rays of light in geometric optics correspond to photons in quantum mechanics and also correspond to energy flux in physical optics, as known. On the principle of quantum mechanics, the conclusion could also be reached that the energy flux is the key factor to make the matter change.

Formation of fringes in the photographic emulsion by standing-waves is also inconsistent with the principle of geometric optics as shown in the section 2.1.

In this paper, Equation (8) or (9) for the interference between two light beams has been derived by means of Energy Flux in Interference, which can be used to calculate the energy flux densities in interference fringes and standing waves. The equations suggest that both electric-field vector and magnetic-field vector are in phase with each other at the maxima of interference fringes of two light beams. The in-phase attribution of interference fringes is just different from standing-waves in which the magnetic-field is in quadrature with the electric-field. In addition, it has been proved that both the magnetic energy density and the electric energy density in the interference fringes or the standing-waves are equal to each other [APPENDIX C]. It consists with the wave principle that electromagnetic waves can only exist as the electric energy and magnetic energy are converting to each other.

Therefore, they also indicate that the magnetic-field vector acts the same role as the electric-field vector on light interacting with substance, and the fundamental factor of optical interference is electromagnetic energy flux densities rather than electric-field alone. The equations are consistent with the results of the experiments and the principles of electromagnetic waves and quantum mechanics and geometric optics.

In summary, the intensity of optical interference fringes should be the energy flux density of $\left|\vec{E}\times\vec{H}\right|$, not electric-field energy density of $\left|E\right|^2$.

## 6 AMAZING DEDUCTION

### (1) Faster-than-light propagating of electromagnetic energy flux in interference field

There is another way to understand the Interference Energy Flux. According to Equation (7), the propagating speed $V_{x-InterferenceFlux}$ of the interference traveling waves is given by

$$V_{x-InterferenceFlux} = \frac{\omega}{k\cos\frac{\theta}{2}} = \frac{V}{\cos\frac{\theta}{2}} = \frac{1}{\sqrt{\varepsilon\mu}\cos\frac{\theta}{2}}.$$

By the speed $V_{x-InterferenceFlux}$, it can be used for easy explaining that the electromagnetic energy density $2a^2\cos^2(\theta/2)$ for the interference field corresponds to the energy flux density $2a^2\cos(\theta/2)$ in the direction of x-axis mentioned in section 4.

But it suggests that the propagating speed of electromagnetic energy flux in interference field is faster than light speed. The detail is presented in APPENDIX D.

### (2) Modified probability density function in quantum mechanics

Max Born, one of the authors of PRINCIPLES of OPTICS[5], formulated the now-standard interpretation of the probability density function for $\left|\psi\right|^2$ in the Schrödinger equation, published in July 1926[10].

$\left|\psi\right|^2$ came from $\left|E\right|^2$ in classical wave function. So it might be modified according to the intensity of interference should be equal to $\left|\vec{E}\times\vec{H}\right|$.

**Acknowledgment.** Ms. Wen Wei quantitatively measured the Bragg diffraction and the ±1st order diffraction of the developed Photographic Plates. Dr. GuanHong Gao and Dr. JiQing Chen made the modification of English in this paper. Several apertures in different shapes and sizes were made in courtesy of HangZhou Astronomy Science Company Ltd. Dr. Ye Pu bought me the original Wiener's article[1].

# APPENDIXES

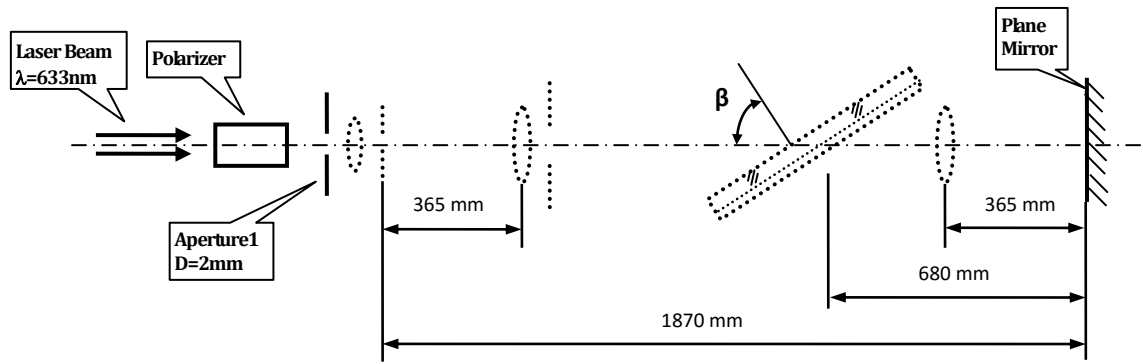

**Fig.A.1. The first step on adjustment for exact optical standing waves (the dotted ones were removed out temporarily)**

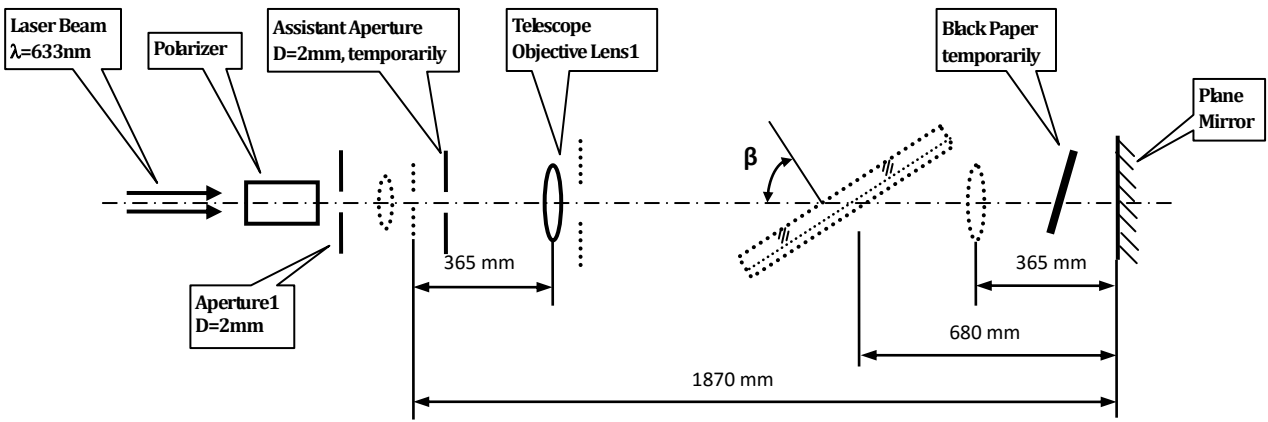

**Fig.A.2. The second step on adjustment for exact optical standing waves (the dotted ones were removed out temporarily)**

# APPENDIX A: Adjustment to build exact optical standing waves

It needs accurate adjustments to build standing waves. The accurate adjustments can ensure to form exactly parallel light beams without optical aberration exactly in the opposite directions in which the incident and reflected light propagate respectively.

The accurate coaxiality of optical parts is the primary objective of the adjustment steps because it determines how much the optical aberration is in the parallel beams. The principal requirement was to place every optical part in the laser beam coming directly from the laser device with a beam diameter of 2~4mm. The beam worked as an optical axis.

Firstly for the accurate experiment setup to be made easily, every optical part was placed as in Fig.4. Subsequently, the process for the adjustment should be carried out as follows.

**The first step:** to set up an optical axis. After removing the Microscope Objective Lens and Pinhole, Telescope Objective Lenses1 and 2, the Aperture2 in diameter of 25mm, and the Photographic Plate, which were all dotted as in Fig.A.1, from the optical path, Aperture1 in diameter of about 2mm was placed in the path behind the Polarizer. The azimuth of the Plane Mirror was adjusted to make the reflected beam back onto the centre of Aperture1. The Plane Mirror should never be touched before the experiment was completed because the light beam had already been set as an optical axis.

**The second step:** to place Telescope Objective Lens1. Telescope Objective Lens1 and the Assistant Aperture were placed in the optical path as in Fig.A.2. The position of the lens was adjusted to make the beam reflected by the Plane Mirror onto the centre of the Assistant Aperture, and then a piece of Black Paper was temporarily placed in the path as a shield from the mirror. The azimuth of the lens was adjusted to make the centre of the interference rings reflected by the lens onto the centre of the Assistant Aperture. Telescope Objective Lens1 was in the optical axis after the second step.

**The third step:** to place the Microscope Objective Lens and the Pinhole with a clear aperture of 5μm in the optical path as in Fig.A.3. Parallel Glass Plate1 with precision of 1 arc-second and Parallel Glass Plate2 only for weakening the laser beam to the Collimator were placed as in Fig.A.3. The quality of the expanded parallel beam was detected by a Collimation Tester with a diameter of 50.8mm placed between Telescope Objective Lens1 and Parallel Glass Plate1 according to Shearing Interferometry. The Collimation Tester, not illustrated in Fig.A.3, was a high-quality optical window with two extremely flat optically-polished uncoated surfaces with a slight wedge (4.4 arc-seconds) between the two surfaces. The distance between the Pinhole and Telescope Objective Lens1 was adjusted according to the Collimation Tester. The azimuth of the Microscope Objective Lens was adjusted until the centre of the interference rings reflected by the Lens was just on the centre of Aperture1. The Pinhole along with its image reflected from the Plane Mirror could be watched in the Collimator. The position of the Microscope Objective Lens with the Pinhole was adjusted in order to place the both images over each other in the Collimator. Once those were done, the positions of the Microscope Objective Lens and the Pinhole were in the optical axis. The crosshairs of the Collimator was regulated to record the position of the images for the following steps.

**The fourth step:** to place Telescope Objective Lens2. After temporarily removing both the Microscope Objective Lens and the Pinhole from the optical path, Telescope Objective Lens2 and the Assistant Aperture were placed in the path as in Fig.A.4. In order to place Telescope Objective Lens2 in the optical axis, the adjustment process was the same as the second step (to place Telescope Objective Lens1). Pay extra attention to keeping the distance between Telescope Objective Lens2 and the Assistant Aperture longer than 650mm, as a failure experience in repeating the experiments which had been successful a half year ago, showed that a less distance made successful number decline severely.

**The fifth step:** to place back the Microscope Objective Lens and the Pinhole. After removing the Assistant Aperture, the Microscope Objective Lens and the Pinhole were placed back as in Fig.A.3, according to the crosshairs of the Collimator that had recorded the position of the Pinhole at the third step.

**The sixth step:** to place a Photographic Plate. After removing Parallel Glass Plate1 and the Black Paper out of the optical path in Fig.A.4, Aperture2 with a diameter of 25mm was fixed as in Fig.4. Then a Photographic Plate was placed, which was not developed and could be used repeatedly, in the proper position and planned azimuth, and its emulsion was faced to the Plane Mirror in the same way as in Wiener's experiment. Telescope Objective Lens2 was then translated perpendicularly to the optical axis in the horizontal plane or the paper plane of Fig.4 to drive the both light spots to superimpose on the emulsion of the Photographic Plate. The both light spots were originated from the Pinhole and reflected from the Plane Mirror respectively.

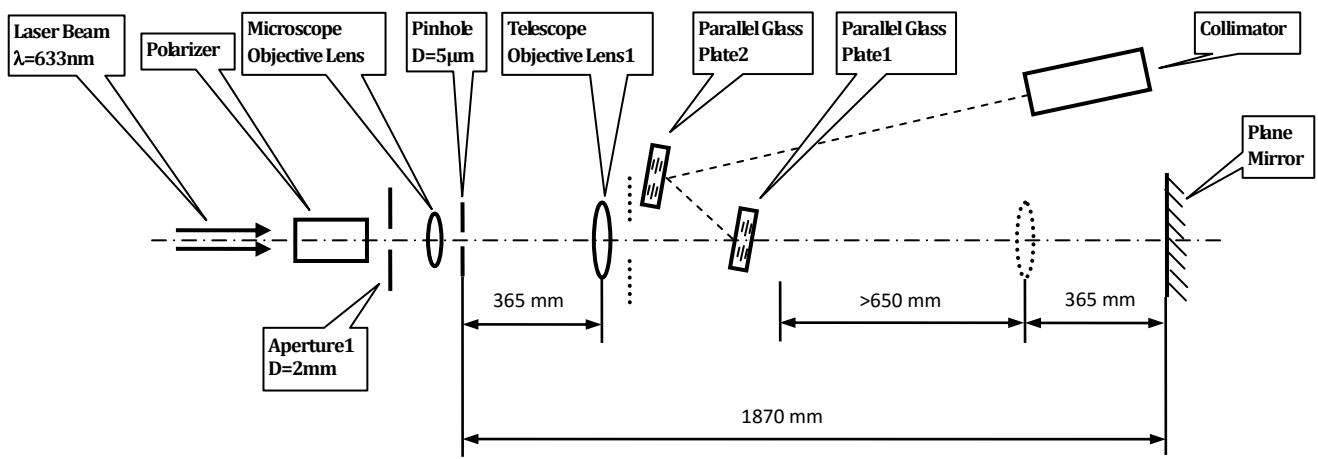

**Fig.A.3. The third step on adjustment for exact optical standing waves (the dotted ones were removed out temporarily and the Collimation Tester were not illustrated)**

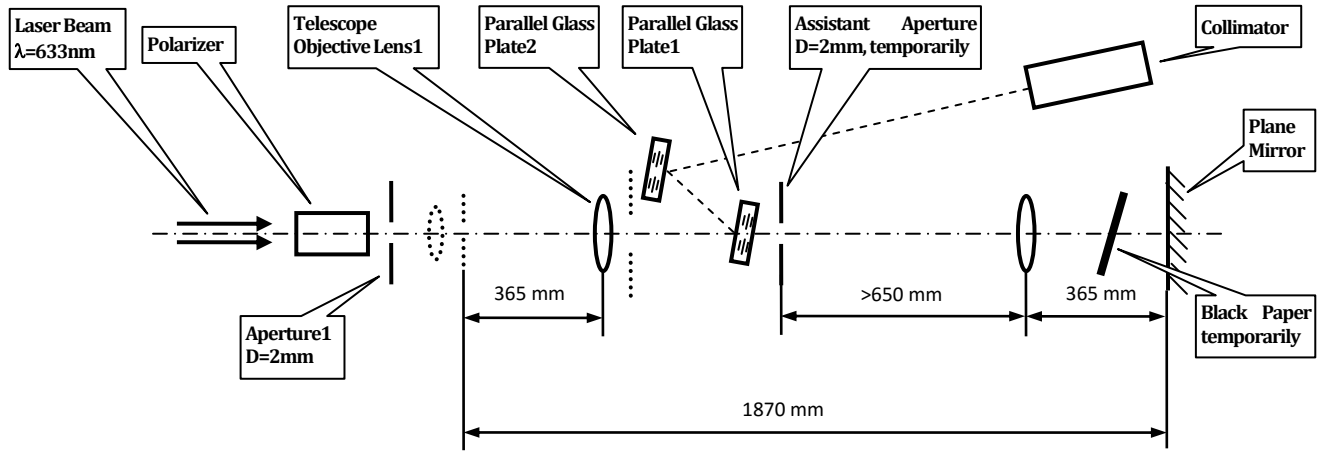

**Fig.A.4. The fourth step on adjustment for exact optical standing waves (the dotted ones were removed out temporarily)**

The mounting to fix the Photographic Plate could be operated in dark for any unexposed Photographic Plate. The Photographic Plate was made in TianJing TianGan Photographic Materials Company Ltd., called TRIRING Holographic Photographic Plate of Model I designed for sensitive wavelength of 633nm. Because the glass plate used as the substrate of the emulsion was not made of optical glass with uniform refractive index, it might cause optical aberration. And its typical thickness of 1mm was too thin to ensure the surface of emulsion flat enough. The photographic plates used in the experiments were not bought directly in the market. In fact, the optical glass plates made of float glass in thickness of 3mm with customized shape of a rectangle of 90mm×240mm as the company's normal size were coated with emulsion in the company. So the Photographic Plates in the experiments were made of optical glass substrates in thickness of 3mm. The parallelity of a Photographic Plate could be detected by means of equal-thickness interference between the surfaces of both glass and emulsion while the plate was illuminated by a parallel light beam. The interference fringe spacing of a normal photographic plate was about 0.8mm, much less than average 4mm spacing of a Photographic Plate made of thicker optical glass. That meant the flatness of the Photographic Plates made of thicker optical glass was much better, and consequently success number of the experiments was much more than those from market.

## APPENDIX B: The Results and Analysis for the Photographs When Exposed at the Incidence Angle of 2.5°

The diffraction results of the developed photographs at 2.5° under a thin laser beam with light power of about 3mW are summarized in Table B. Because more than one interference patterns were on the photographs, they disturbed the Bragg diffraction formed by the standing waves. Thus, the chances for photographs to show no Bragg diffraction were determined by the photograph's parallelity and thickness instead of parallelity and flatness. The parallelity of these Photographs is also listed in the table as the number of the equal-thickness fringes.

There were six interference patterns on the photographs in this case. In the detection of a developed Photographic Plate as in Fig.6, we could get the diffracted light from the developed photographs in three directions by using a thin laser beam irradiating on it:

**i.** In the anti-incidence direction, the Bragg diffraction from Fringe ①② including the diffraction light beams from Fringe ②③ and Fringe ①④ could be found. As the detecting thin laser beam went in the way of Beam ①, the diffraction light beam from Fringe ①② would travel in the way of Beam ②, namely Bragg diffraction. Because the inner surface of the emulsion made the detecting thin laser beam be reflected in the way of Beam ③ and the reflected would be the incident beam on Fringe②③ as in Fig.5, the diffraction light beam from Fringe ②③ would travel in the way of Beam ② which was just the direction of Bragg diffraction. Because the surfaces of the sides of a photograph were almost parallel, Fringe ①④ and Fringe ②③ were very similar to each other on spacing and direction of the fringes. Thus, Fringe ①④ had the same effect as Fringe②③.

**ii.** In the direction of the ±1st order diffraction, the diffraction light beams from Fringe ①④, Fringe ②③ and Fringe ③④ should be found. However, Fringe ③④ was much weaker than the others as both Beam ③ and Beam ④ were reflected respectively from the inner surfaces of the photographic plates.

**iii.** In the direction of the reflection from the surfaces of the developed Photographic Plates, the diffraction light beams from Fringe ①③ and Fringe ②④ should be found. But they were not detected in the direction because they could not be separated from the reflection.

According to Table B, the results for the angle of 2.5° are:

**i.** The diffraction in the anti-incidence direction were not found on 11 photographs whose two surfaces were very parallel based on their fringes of equal thickness, also without the diffraction of the ±1st order, or with very weak diffraction of the ±1st order.

Though Fringe ②③ was in the identical fringe spacing and fringe direction with Fringe ①④ when the surfaces of two sides of a photograph were parallel to each other, their initial phase angles or beginning points of the fringes on the emulsion were different from each other. When the difference of their initial phase angles on the emulsion due to the thickness of the photographic plate was 180°, the maxima of Fringe ②③ were just at the minima of Fringe ①④. It suggested that both fringes disappeared when they overlapped each other in the way of the maxima of a fringe on the minima of the other fringe. In this case, the diffraction from Fringe ②③ and Fringe ①④ could not be found in their diffraction directions. It also explained the phenomenon of no the ±1st order diffraction on 3 photographs.

**Table B. Statistics of the results of photographs exposed at 2.5° and their parallelity listed as the equal-thickness fringes' number**

| | | Number of photos | Number of fringes; Distribution |
|---|---|---|---|
| No Bragg diffraction | without the ±1st order diffraction | 3 | less than 4 |
| | with the ±1st order diffraction: 5~49nW | 8 | |
| Bragg diffraction: 6~20nW | with the ±1st order diffraction: 4~342nW | 32 | about 4 |
| Bragg diffraction: 21~112nW | with the ±1st order diffraction: 6~1630nW | 69 | much more than 4 |
| Total of the photographs | | 112 | |

If both fringes did not disappear at all, the diffraction from Fringe ②③ and Fringe ①④ might be too weak to see in the anti-incidence direction as the reflecting incidence light beam was the reflection of the laser beam from the inner surface of the emulsion. The reflecting incidence was about 4% of the incidence light in intensity. But the weak ±1st order diffractions could been seen as the incidence light beam was thin laser beam. Another probable reason for no diffraction in the anti-incidence direction was that the interference of the Bragg diffraction from the fringe ①② and the diffractions from Fringe ②③ and Fringe ①④ was just to be zero.

The conclusion could also be drawn that the optical standing waves were not able to make any interference fringes in the emulsion based on the fact that both Bragg diffraction from Fringe ①② and the diffraction of the ±1st order from Fringe ②③ or Fringe ①④ could not be found simultaneously even though on 3 photographs.

Furthermore, even though the area of a photograph exposed at the angle of 2.5° (the less the area was, the more the number of perfectly parallel two surfaces of a photograph was.) was two times less than at the angle of 57.5°, the 3 photographs without both the Bragg diffraction and the ±1st order diffraction at the angle of 2.5° was much less than 11 at the angle of 57.5°. The reason was that the phenomenon without both the Bragg diffraction and the ±1st order diffraction in this case required the following two conditions. One was a perfect parallelity of two surfaces of a photograph. The other was the proper thickness of the photograph.

**ii.** On the most of photographs, the Bragg diffraction and the ±1st order diffraction at the angles of ±2×2.5° could be found. This would happen when the two surfaces of photographs were not very parallel based on the measurement of their fringes of equal thickness or the thickness of the photographs were not proper in the way that Fringe ②③ and Fringe ①④ cancelled each other out.

The ±1st order diffraction came also from Fringe ②③ which showed up definitely but was never mentioned in Wiener's experiment. Actually it was Fringe ②③ that appeared in Wiener's experiments.

# APPENDIX C: Derivation for Energy Flux in Interference Field and No Energy Flux in Standing Waves

## C.1 The General Description of the Two Beams in Fig.7

If for the two beams the two electric-field vectors are denoted as $\vec{E_1}$ and $\vec{E_2}$, the two magnetic field vectors as $\vec{H_1}$ and $\vec{H_2}$, as shown in Fig.7, they could be expressed by

$$\vec{E_1} = \vec{E_{10}} \cdot \exp[ik(x\cos\frac{\theta}{2} - y\sin\frac{\theta}{2})] \quad , \quad \text{or}$$

$$\vec{E_1} = \vec{E_{10}} \cdot \cos[\omega t - k(x\cos\frac{\theta}{2} - y\sin\frac{\theta}{2})] \quad (C1)$$

$$\vec{H_1} = \vec{H_{10}} \cdot \exp[ik(x\cos\frac{\theta}{2} - y\sin\frac{\theta}{2})] \quad , \quad \text{or}$$

$$\vec{H_1} = \vec{H_{10}} \cdot \cos[\omega t - k(x\cos\frac{\theta}{2} - y\sin\frac{\theta}{2})] \quad (C2)$$

$$\vec{E_2} = \vec{E_{20}} \cdot \exp[ik(x\cos\frac{\theta}{2} + y\sin\frac{\theta}{2})] \quad , \quad \text{or}$$

$$\vec{E_2} = \vec{E_{20}} \cdot \cos[\omega t - k(x\cos\frac{\theta}{2} + y\sin\frac{\theta}{2})] \quad (C3)$$

$$\vec{H_2} = \vec{H_{20}} \cdot \exp[ik(x\cos\frac{\theta}{2} + y\sin\frac{\theta}{2})] \quad , \quad \text{or}$$

$$\vec{H_2} = \vec{H_{20}} \cdot \cos[\omega t - k(x\cos\frac{\theta}{2} + y\sin\frac{\theta}{2})] \quad (C4)$$

where, $|\vec{E_{10}}| = |\vec{E_{20}}| = a$, $|\vec{H_{10}}| = |\vec{H_{20}}| = b$, $\omega = 2\pi\nu$, $k = 2\pi/\lambda$, $\lambda$ is the wavelength of the light beams, and $\nu$ is the frequency of the light.

For one beam, e.g. the first beam, the energy density of electric field is $w_{e1} = \frac{\varepsilon}{2}|\vec{E_1}|^2$, and its time-averaged value is $\langle w_{e1} \rangle_T = \frac{\varepsilon}{4}a^2$.

The energy density of magnetic field is $w_{m1} = \frac{\mu}{2}|\vec{H_1}|^2$, and its time-averaged value is $\langle w_{m1} \rangle_T = \frac{\mu}{4}b^2$.

According to the electromagnetic equation:
$$\sqrt{\varepsilon}a = \sqrt{\mu}b \quad (C5)$$

Where ε is permittivity and μ is magnetic permeability, the time-averaged value of the energy density of electromagnetic field is

$$\langle w_{em1} \rangle_T = \langle w_{e1} \rangle_T + \langle w_{m1} \rangle_T = \frac{\varepsilon}{2}a^2$$

The energy flux density or Poynting vector for the first beam is $\vec{S_1} = \vec{E_1} \times \vec{H_1}$. Its time-averaged value of energy flux density is

$$\langle |\vec{S_1}| \rangle_T = \frac{1}{2}ab = \frac{1}{2}\sqrt{\frac{\varepsilon}{\mu}}a^2 = \langle w_{em1} \rangle_T \cdot V \quad , \text{where } V \text{ is the}$$

light velocity, and $V = 1/\sqrt{\varepsilon\mu}$.

## C.2 The Relation Between Resolutions of Poynting Vector and the Magnetic Field Vector in Fig.7

For the two beams, the energy flux densities or Poynting vectors are $\vec{S_1}$ and $\vec{S_2}$ respectively described by

$$\vec{S_1} = \vec{E_1} \times \vec{H_1} \quad (C6)$$

and

$$\vec{S_2} = \vec{E_2} \times \vec{H_2} \quad (C7)$$

If resolving the magnetic field vector $\overrightarrow{H_2}$ into the components in x-axis and y-axis as in Fig.7, we can get $\overrightarrow{H_2} = H_{2x}\overrightarrow{x_0} + H_{2y}\overrightarrow{y_0}$, where, $\overrightarrow{x_0}$ and $\overrightarrow{y_0}$ are unit vectors along x-axis and y-axis respectively, then its energy flux density is

$$\overrightarrow{S_2} = \overrightarrow{E_2} \times \overrightarrow{H_2} = E_{2z}\overrightarrow{z_0} \times (H_{2x}\overrightarrow{x_0} + H_{2y}\overrightarrow{y_0}) \quad (C8)$$

And if we resolve $\overrightarrow{S_2}$ into the components in x-axis and y-axis too, then

$$\overrightarrow{S_2} = S_{2x}\overrightarrow{x_0} + S_{2y}\overrightarrow{y_0} \quad (C9)$$

By comparing Equation (C9) with Equation (C8), we have

$$S_{2x}\overrightarrow{x_0} = E_{2z}H_{2y}(\overrightarrow{z_0} \times \overrightarrow{y_0}) = -E_{2z}H_{2y}\overrightarrow{x_0} \quad (C10)$$

$$S_{2y}\overrightarrow{y_0} = E_{2z}H_{2x}(\overrightarrow{z_0} \times \overrightarrow{x_0}) = E_{2z}H_{2x}\overrightarrow{y_0} \quad (C11)$$

$S_{2x}$ is the energy flux density along x-axis described by

$$S_{2x} = ab\cos\frac{\theta}{2}\cdot\cos^2[\omega t - k(x\cos\frac{\theta}{2} + y\sin\frac{\theta}{2})] \quad (C12)$$

Its time-averaged value is

$$\langle S_{2x}\rangle_T = \frac{ab}{2}\cos\frac{\theta}{2} \quad (C13)$$

Similarly, the time-averaged value of the component along y-axis is,

$$\langle S_{2y}\rangle_T = \frac{ab}{2}\sin\frac{\theta}{2} \quad (C14)$$

If observing any unit area on the beam transverse section, e.g. the area of SS as in Fig.7, $|\overrightarrow{S_2}|$ is the energy flux passing through the unit area of SS. And $S_{2x}$ is the energy flux density passing through the area PP as in Fig.7, so the time-averaged value of the energy flux through the area PP is

$$\langle EF_{PP-2x}\rangle_T = \langle S_{2x}\rangle_T \cdot SS \cdot \cos\frac{\theta}{2} = \frac{ab}{2}\cos^2\frac{\theta}{2} \quad (C15)$$

Similarly, the energy flux through the area QQ is

$$\langle EF_{QQ-2y}\rangle_T = \langle S_{2y}\rangle_T \cdot SS \cdot \sin\frac{\theta}{2} = \frac{ab}{2}\sin^2\frac{\theta}{2} \quad (C16)$$

The sum of the two time-averaged values of the energy flux is

$$\langle EF_{PP-2x}\rangle_T + \langle EF_{QQ-2y}\rangle_T = \frac{ab}{2} \quad (C17)$$

That is the time-averaged value of the energy flux $\overrightarrow{S_2}$ passing through the unit area SS. The result indicates that the way to resolve $\overrightarrow{S_2}$ and $\overrightarrow{H_2}$ is consistent with conservation of energy.

**C.3 Interference Energy Flux Density and Existence of Standing-wave Component**

By resolving the magnetic-fields of the two beams into the components, $H_{1x}$ and $H_{2x}$ in x-axis, and $H_{1y}$ and $H_{2y}$ in y-axis respectively, we have

$$H_{1x} = -b\sin\frac{\theta}{2}\cdot\exp[ik(x\cos\frac{\theta}{2} - y\sin\frac{\theta}{2})] \quad (C18)$$

$$H_{1y} = -b\cos\frac{\theta}{2}\cdot\exp[ik(x\cos\frac{\theta}{2} - y\sin\frac{\theta}{2})] \quad (C19)$$

$$H_{2x} = b\sin\frac{\theta}{2}\cdot\exp[ik(x\cos\frac{\theta}{2} + y\sin\frac{\theta}{2})] \quad (C20)$$

$$H_{2y} = -b\cos\frac{\theta}{2}\cdot\exp[ik(x\cos\frac{\theta}{2} + y\sin\frac{\theta}{2})] \quad (C21)$$

The electric-fields polarizing along z-axis are

$$E_{1z} = a\cdot\exp[ik(x\cos\frac{\theta}{2} - y\sin\frac{\theta}{2})] \quad (C22)$$

$$E_{2z} = a\cdot\exp[ik(x\cos\frac{\theta}{2} + y\sin\frac{\theta}{2})] \quad (C23)$$

In the intersecting area of the beams, the superposition of the magnetic-fields along x-axis is,

$$H_x = H_{1x} + H_{2x} \quad (C24)$$

So,

$$H_x = 2b\sin\frac{\theta}{2}\sin(ky\sin\frac{\theta}{2})\cdot i\cdot\exp[ikx\cos\frac{\theta}{2}] \quad (C25)$$

Similarly,

$$H_y = -2b\cos\frac{\theta}{2}\cos(ky\sin\frac{\theta}{2})\cdot\exp(ikx\cos\frac{\theta}{2}) \quad (C26)$$

and

$$E_z = 2a\cos(ky\sin\frac{\theta}{2})\cdot\exp[ikx\cos\frac{\theta}{2}] \quad (C27)$$

If Equations (C25), (C26) and (C27) are expressed in time domain, then,

$$H_x = 2b\sin\frac{\theta}{2}\sin(ky\sin\frac{\theta}{2})\cdot\sin(\omega t - kx\cos\frac{\theta}{2}) \quad (C28)$$

$$H_y = -2b\cos\frac{\theta}{2}\cos(ky\sin\frac{\theta}{2})\cdot\cos(\omega t - kx\cos\frac{\theta}{2}) \quad (C29)$$

$$E_z = 2a\cos(ky\sin\frac{\theta}{2})\cdot\cos(\omega t - kx\cos\frac{\theta}{2}) \quad (C30)$$

Equations (C25) and (C27) or (C28) and (C30) indicate the magnetic-field component Hx is in quadrature with the electric-field vector Ez and then indicate existence of the standing-wave component in the interference field. Its instantaneous energy flux density along y-axis is,

$$S_y = E_z \cdot H_x \text{, then}$$

$$S_y = ab\sin\frac{\theta}{2}\sin(2ky\sin\frac{\theta}{2})\cdot\sin[2(\omega t - kx\cos\frac{\theta}{2})] \quad (C31)$$

Then its time-averaged value of the energy flux density along y-axis is

$$\langle S_y\rangle_T = 0 \quad (C32)$$

The corresponding instantaneous energy flux density through yz-plane along x-axis, called Interference Energy Flux Density, is,

$$S_x = -E_z \cdot H_y \text{, then}$$

$$S_x = 4ab\cos\frac{\theta}{2}\cos^2(ky\sin\frac{\theta}{2})\cdot\cos^2(\omega t - kx\cos\frac{\theta}{2}) \quad (C33)$$

where, the component Hy of the magnetic-field is in phase with the electric-field vector Ez. Its time-averaged value of the energy flux density along x-axis is,

$$\langle S_x\rangle_T = 2ab\cos\frac{\theta}{2}\cos^2(ky\sin\frac{\theta}{2}) \text{, or}$$

$$\langle S_x\rangle_T = ab\cos\frac{\theta}{2}[1 + \cos(2ky\sin\frac{\theta}{2})] \quad (C34)$$

and according to Equation (C5), we have

$$\langle S_x\rangle_T = \sqrt{\frac{\varepsilon}{\mu}}a^2\cos\frac{\theta}{2}[1 + \cos(2ky\sin\frac{\theta}{2})] \quad (C35)$$

The space-time-averaged value of the energy flux density along x-axis is

$$\langle S_x\rangle_{TS} = \left\langle 2ab\cos\frac{\theta}{2}\cos^2(ky\sin\frac{\theta}{2})\right\rangle_S = ab\cos\frac{\theta}{2} \quad (C36)$$

The total energy flux through and perpendicular to the yz-plane, or along x-axis, is,

$$\langle EF_x \rangle_T = \langle S_x \rangle_{TS} \cdot DD = \langle S_x \rangle_{TS} \cdot A / \cos\frac{\theta}{2} = abA \quad (C37)$$

All energy flux of $\vec{S_1}$ and $\vec{S_2}$ has entered into the intersecting area according to Equation (C37). It also means that the analysis above is consistent with energy conservation.

The interference result of the two parallel light beams is shown in Equation (C34) or (C35). It misses the ratio of $\cos(\theta/2)$ in the general equation derived only from the electric-fields. Equation (C34) or (C35) is the distribution of the energy flux density in the interference field including electric and magnetic field rather than electric-field energy density alone. Because the phase distribution in Equation (C34) or (C35) is identical to the interference equation by means of electric-field vectors, we can, of course, get the same typical characters of interference fringes as follows.

(1) The fringe direction is parallel to xz-plane, namely, parallel to the bisector of the angle θ between the two beams in the xy-plane.

(2) The fringe spacing is $\Delta y = \lambda / 2\sin\frac{\theta}{2}$.

When $2ky\sin\frac{\theta}{2} = 2\pi m$ ( m=0, 1, 2...), the energy flux density becomes maximum.

(3) The intensity maximum of interference fringes on the yz-plane is 4 times as much as the intensity of only one beam:

$$2ab\cos\frac{\theta}{2} / \langle S_{2x} \rangle_T = 4 .$$

**C.4 The Relation between Interference Energy Flux and the Electromagnetic Energy Densities**

If the two light beams are traveling in a no-loss medium as in Fig.7, at any point in the intersecting area of the beams, the electromagnetic energy flux entering to the point is equal to the increase rate of electromagnetic energy density according to Poynting's theorem, and in this case we have

$$-\nabla \cdot \overrightarrow{S_{1+2}} = \frac{\partial}{\partial t}\left(\frac{\varepsilon}{2}|\vec{E}|^2 + \frac{\mu}{2}|\vec{H}|^2\right) \quad (C38)$$

that is

$$-\nabla \cdot \left(S_x \cdot \vec{x_0} + S_y \cdot \vec{y_0}\right) = \frac{\partial}{\partial t}\left[\frac{\varepsilon}{2}E_z^2 + \frac{\mu}{2}\left(H_x^2 + H_y^2\right)\right] \quad (C39)$$

or

$$-\frac{\partial S_x}{\partial x} - \frac{\partial S_y}{\partial y} = \frac{\partial}{\partial t}\left[\frac{\varepsilon}{2}E_z^2 \cdot \sin^2\left(\frac{\theta}{2}\right) + \frac{\mu}{2}H_x^2\right]$$
$$+ \frac{\partial}{\partial t}\left[\frac{\varepsilon}{2}E_z^2 \cdot \cos^2\left(\frac{\theta}{2}\right) + \frac{\mu}{2}H_y^2\right] \quad (C40)$$

After calculating according to Equations (C28), (C29), (C30), (C31), (C33), and noticing that the equation above holds true for any angle of θ, we have

$$-\frac{\partial S_y}{\partial y} = \frac{\partial}{\partial t}\left[\frac{\varepsilon}{2}E_z^2 \cdot \sin^2\left(\frac{\theta}{2}\right) + \frac{\mu}{2}H_x^2\right] \quad (C41)$$

and $-\frac{\partial S_x}{\partial x} = \frac{\partial}{\partial t}\left[\frac{\varepsilon}{2}E_z^2 \cdot \cos^2\left(\frac{\theta}{2}\right) + \frac{\mu}{2}H_y^2\right] \quad (C42)$

They indicate that the energy flux density Sy of the standing-wave is formed by the electromagnetic energy densities of $\left[\frac{\varepsilon}{2}E_z^2 \cdot \sin^2\left(\frac{\theta}{2}\right) + \frac{\mu}{2}H_x^2\right]$, and the energy flux density $S_x$ of the interference is formed by $\left[\frac{\varepsilon}{2}E_z^2 \cdot \cos^2\left(\frac{\theta}{2}\right) + \frac{\mu}{2}H_y^2\right]$. And according to Equation (C5), we have

$$\left\langle \frac{\varepsilon}{2}E_z^2 \cdot \sin^2\left(\frac{\theta}{2}\right) \right\rangle_T = \left\langle \frac{\mu}{2}H_x^2 \right\rangle_T \quad (C43)$$

$$\left\langle \frac{\varepsilon}{2}E_z^2 \cdot \cos^2\left(\frac{\theta}{2}\right) \right\rangle_T = \left\langle \frac{\mu}{2}H_y^2 \right\rangle_T \quad (C44)$$

Equations (C43) and (C44) indicate that both the electric energy density and the magnetic energy density in standing-waves or interference-field are equal to each other. It consists with the wave principle that electromagnetic waves can only exist on the electric energy and magnetic energy converting to each other.

To iterate the above conclusion, it is the energy flux density in Equation (C34) or (C35) that forms the interference fringes. The energy flux left its traces in the emulsion of a Photographic Plate or on a screen in the same way as streams would do on a field. It is emphasized that the component of magnetic-field vector which works with the electric-field vector and forms the interference fringes has the same phase as the electric-field vector in the energy flux density of the interference, and both the magnetic energy density and the electric energy density that make the interference fringes or the standing-waves are equal to each other. Therefore, they also indicate that the magnetic-field vector acts the same role as the electric-field vector on light interacting with substance, and the fundamental factor of optical interference is electromagnetic energy flux densities rather than electric-field alone. The equations are consistent with the results of the experiments and the principles of electromagnetic waves and quantum mechanics and geometric optics.

## APPENDIX D: Amazing Deduction

There is another way to understand the Interference of Energy Flux and Standing Waves.

In the intersecting area, the energy flux density $\overrightarrow{S_{1+2}}$ of the two beams as in Fig.7 consists of two components. The component $S_y$ is for the standing waves along y-axis and the component $S_x$ along x-axis is for the interference traveling waves.

The average electric energy density is the sum of the electric energy densities of the two beams: $\langle w_e \rangle_T = \frac{\varepsilon}{2}a^2$, and the average magnetic energy density is the sum of the magnetic energy densities of the two beams: $\langle w_m \rangle_T = \frac{\mu}{2}b^2$.

$S_y$ is expressed by Equation (C31):
$$S_y = ab\sin\frac{\theta}{2}\sin(2ky\sin\frac{\theta}{2}) \cdot \sin[2(\omega t - kx\cos\frac{\theta}{2})]$$

Where, there is no propagating term along y-axis in it. It means that $S_y$, the component of the standing waves does not propagate.

The energy flux density of the interference traveling waves, $S_x$ is expressed by Equation (C33):
$S_x = -E_z \cdot H_y$ ,then,
$$S_x = 4ab\cos\frac{\theta}{2}\cos^2(ky\sin\frac{\theta}{2}) \cdot \cos^2(\omega t - kx\cos\frac{\theta}{2})$$

Where , $E_z$ is the electric-field of the interference traveling waves along z-axis given by Equation (C30):

$$E_z = 2a\cos(ky\sin\frac{\theta}{2})\cdot\cos(\omega t - kx\cos\frac{\theta}{2}).$$

and $H_y$ is the magnetic-field of the interference traveling waves along y-axis given by Equation (C29):

$$H_y = -2b\cos\frac{\theta}{2}\cos(ky\sin\frac{\theta}{2})\cdot\cos(\omega t - kx\cos\frac{\theta}{2}).$$

The propagating term of the interference traveling waves along x-axis in the equations is $\cos(\omega t - kx\cos\frac{\theta}{2})$.

Therefore, the propagating speed $V_{x-InterferenceFlux}$ of the interference traveling waves is given by

$$V_{x-InterferenceFlux} = \frac{\omega}{k\cos\frac{\theta}{2}} = \frac{V}{\cos\frac{\theta}{2}} = \frac{1}{\sqrt{\varepsilon\mu}\cos\frac{\theta}{2}}.$$

Because only the components, $H_{1y}$ and $H_{2y}$ in the magnetic-field in the interference energy flux are traveling along x-axis, the average magnetic energy density in the interference is

$$\langle w_{m-Interference}\rangle_T = \frac{\mu}{2}b^2\cos^2\frac{\theta}{2}.$$

According to Equation (C44) or the wave principle that electromagnetic waves can be traveling only by means of the electric energy and magnetic energy converting to each other, there must be the same amount of energy of the electric-field as the magnetic-field in the interference. So the average electric energy density in the interference is given by

$$\langle w_{e-Interference}\rangle_T = \frac{\varepsilon}{2}a^2\cos^2\frac{\theta}{2}.$$

Therefore, the whole average electromagnetic energy density in the interference is

$$\langle w_{em-Interference}\rangle_T = \langle w_{e-Interference}\rangle_T + \langle w_{m-Interference}\rangle_T,$$

then,

$$\langle w_{em-Interference}\rangle_T = \sqrt{\varepsilon\mu}\,ab\cos^2\frac{\theta}{2}$$

and, the space-time-averaged value of the energy flux density in the interference is

$$\langle S_x\rangle_{TS} = \langle w_{em-Interference}\rangle_T \cdot V_{x-InterferenceFlux} = ab\cos\frac{\theta}{2}$$

This result is identical to Equation (C36).

Due to the analysis above, an amazing deduction has been drawn that the propagating speed of electromagnetic energy flux in interference field is faster than light speed.

# UNNECESSARY APPENDIXES

### UNNECESSARY APPENDIX a: the interference fringes between a glass surface of the Glass Plate and a plane mirror

For further explanation, a simple experiment was presented here. As shown in Fig.U.1, the fringes formed by the interference between a glass surface of the Glass Plate and a plane mirror could be observed on the white paper shown in Fig.U.2. A very flat surface of the Glass Plate was faced to the mirror. Fig.U.2 and Fig.U.3 were the photos of the actual setup. The fringes with their visibility of 0.4 on the White Paper in Fig.U.1 were shown in Fig.U.4 and Fig.U.5.

The clear fringes with their visibility of 0.4 could explain that the measured photocurrent shown in Ref. [7-9] was generated by the fringes formed by the interference between beam③ (reflected from the transparent photodiode surface) and beam② (reflected from the mirror).

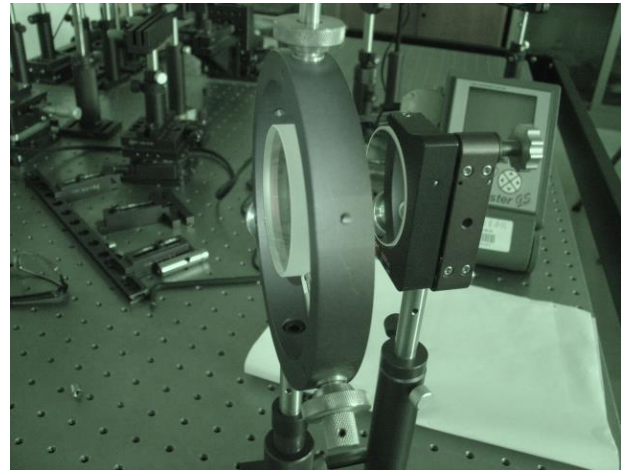

**Fig.U.3 The photo of the actual setup in another view**

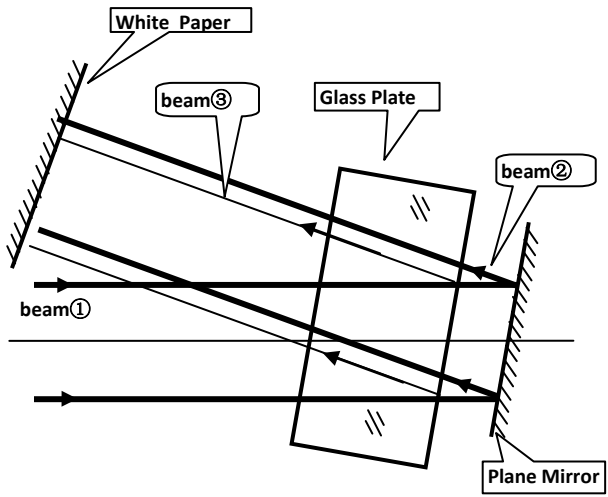

**Fig. U.1 The setup for the fringes formed by the interference between a glass surface and a plane mirror**

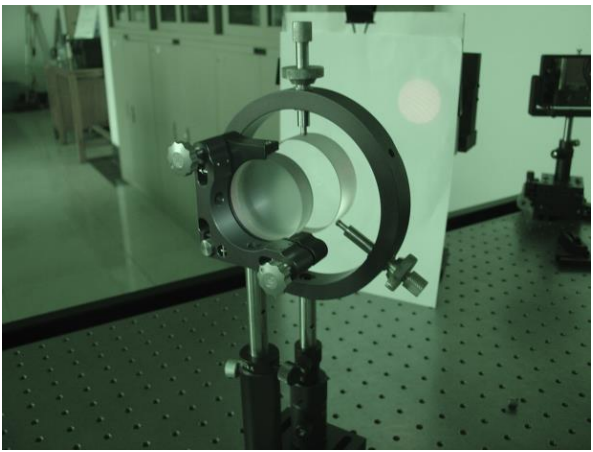

**Fig.U.2 The photo of the actual setup based on Fig.U.1 under normal room light**

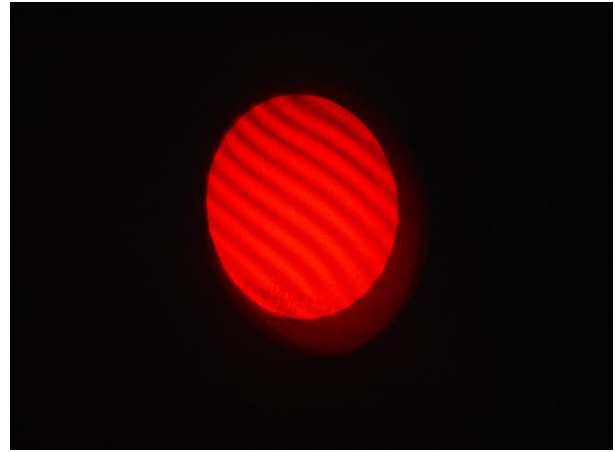

(a) A photo was taken by a digital camera of SONY DSC-V1, with F3.2, 1/30 second without room light.

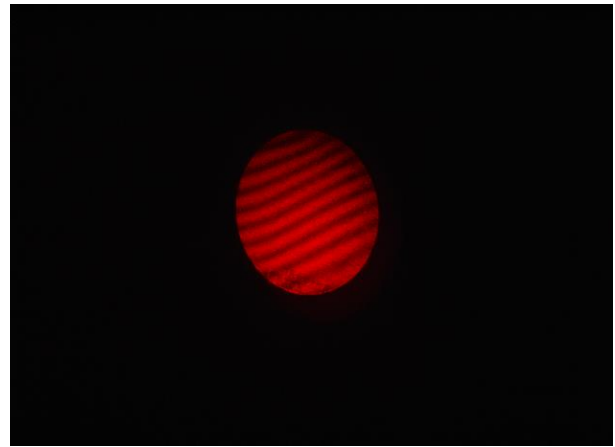

(b) A photo was taken by a digital camera of SONY DSC-V1, with F3.2, 1/60 second without room light.

**Fig.U.4 The photos of the fringes on the White Paper in Fig.U.1**

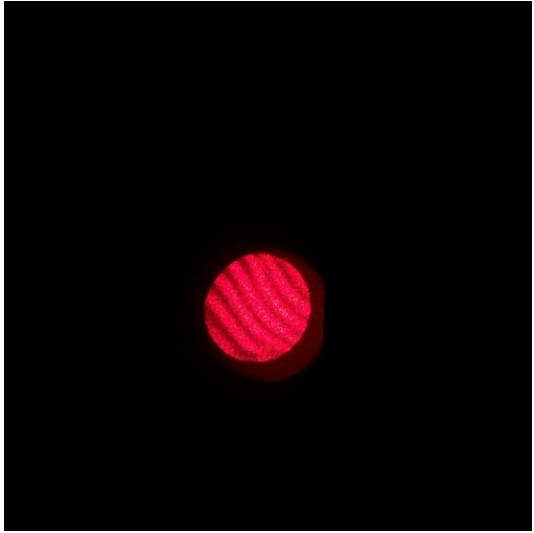 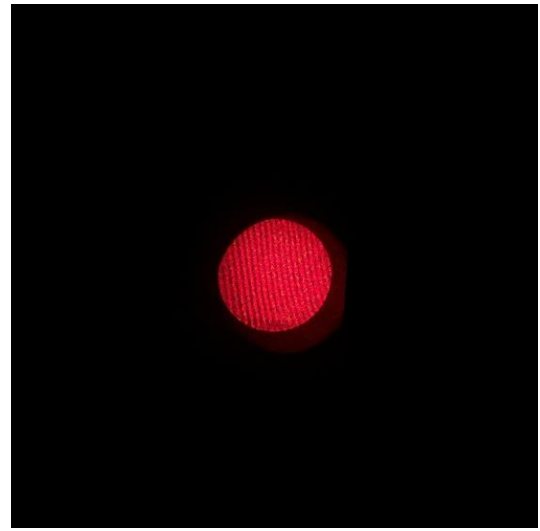

(a) A photo was taken automatically by an iPhone6 camera without room light.

(b) A photo was taken automatically by an iPhone6 camera without room light.

**Fig. U.5 The photos of the other fringes on the White Paper in Fig.U.1**

## UNNECESSARY APPENDIX b: The function of the Telescope Objective Lens2

If the surface of the emulsion was not parallel to the surface of glass plate, the whole Photographic Plate would behave as a glass wedge. In this case as shown in Fig.U.6, the incidence of beam① would turn a little angle against its original direction after passing the glass wedge, and then converge at a point, rather than just exactly at the focus, on the Telescope Objective Lens2's focus plane. After being reflected back by the Plane Mirror, the reflected beam② would go back just in the opposite and parallel direction of the incident beam with a little translation. In the other words, the parallel reflected beam② would encounter the incidence of beam① just oppositely in the emulsion.

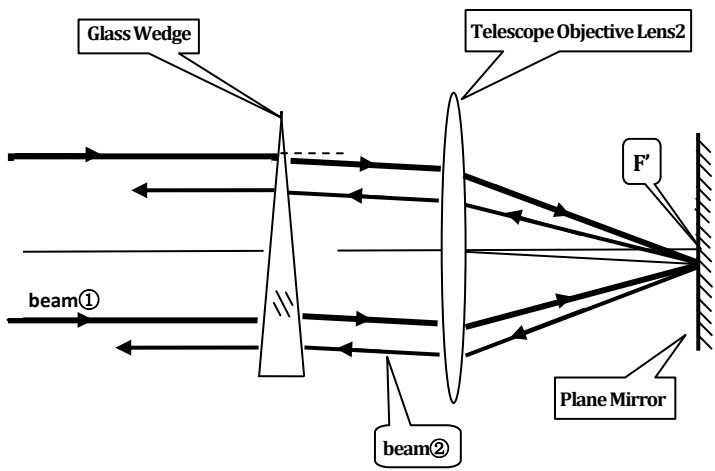

**Fig.U.6 The function of the Telescope Objective Lens2 while the Photographic Plate became a glass wedge.**